\documentclass[aps,prl,showpacs,amsmath,twocolumn,amssymb,superscriptaddress,letterpaper]{revtex4}

\usepackage{graphicx,color}
\usepackage{amssymb}   
\usepackage{amsmath}
\usepackage{epstopdf}
\usepackage{natbib}
\usepackage{hyperref}
\usepackage{bm}

\begin{document}

\preprint{APS/123-QED}

\title{Quantifying Non-Abelian Stability in Majorana Qubits through Rabi Beating Signatures}

\author{Yu Zhang}
\affiliation{School of Physics, MOE Key Laboratory for Non-equilibrium Synthesis and Modulation of Condensed Matter, Xi’an Jiaotong University, Xi’an 710049, China}

\author{Jiayi Chen}
\affiliation{School of Physics, MOE Key Laboratory for Non-equilibrium Synthesis and Modulation of Condensed Matter, Xi’an Jiaotong University, Xi’an 710049, China}

\author{Jie Liu}
\thanks{Corresponding author: jieliuphy@xjtu.edu.cn}
\affiliation{School of Physics, MOE Key Laboratory for Non-equilibrium Synthesis and Modulation of Condensed Matter, Xi’an Jiaotong University, Xi’an 710049, China}
\affiliation{Hefei National Laboratory, Hefei 230088, China}

\author{X. C. Xie}
\thanks{Corresponding author: xcxie@pku.edu.cn}
\affiliation{International Center for Quantum Materials, School of Physics, Peking University, Beijing 100871, China}
\affiliation{Interdisciplinary Center for Theoretical Physics and Information Sciences, Fudan University, Shanghai 200433, China}
\affiliation{Hefei National Laboratory, Hefei 230088, China}

\begin{abstract}
Evaluating the stability of Majorana qubits (MQs) is a central challenge for topological quantum computation. 
Here we propose a simple and experimentally accessible protocol to quantify MQ stability by coupling a quantum dot (QD) to an MQ, which induces Rabi oscillations in the QD charge occupation that can be directly detected using recently developed single-shot readout techniques. 
In realistic systems, deviations from ideal MQ behavior lead to a characteristic beating pattern in the Rabi dynamics. 
We show that the beating frequency scales linearly with these deviations while remaining independent of the base Rabi frequency, thereby providing a direct and quantitative measure of MQ stability. 
Importantly, the beating signature is robust against weak dissipation, and we further demonstrate that the effective model remains quantitatively accurate when benchmarked against a realistic minimal Kitaev chain. 
Our results establish a practical and scalable route for quantitatively characterizing Majorana qubit stability in current experimental platforms.
\end{abstract}


\maketitle



\textit{Introduction.} 
Majorana zero modes (MZMs) in topological superconductors (TS) have garnered significant attention in the condensed matter physics community over the
 past two decades due to their potential for topological quantum computation\cite{kitaev, nayak, Ivanov, alicea3, Fu, sau, fujimoto, sato, alicea2, lut, kou, deng, das1, hao1, Marcus, deng2, perge, Yaz2, PJJ1, PJJ2, QZB1,QZB2}. 
 Unlike conventional fermions or bosons, MZMs obey non-Abelian statistics, whereby braiding operations implement nontrivial unitary transformations on the quantum state, forming the basis of topological quantum computation\cite{kitaev, nayak, Ivanov, alicea3}.
However, the emergence of trivial Andreev bound states (ABSs) in proximitized nanowires poses a critical challenge to realizing topological quantum computation\cite{Jie1, brouwer, Aguado, AguadoN, ABSM, Moore, ChunXiao1, Aguado2, Wimmer, Tewari, 075161, 184520, 035312, 155314, 075416, 054510, 013377}. These ABSs, which can be induced by inhomogeneous potentials or disorder, are difficult to eliminate using existing nanofabrication techniques\cite{ABS1, ABS2, ABS3, ABSM1, ABS4, ABS5, ABS6, ABS7, ABS8, ABS9, Klinovaja, MR1, MR2}. 
Consequently, understanding and mitigating their impact has become a central focus of the field.
 
In general, ABSs can be viewed as two partially separated MZMs, $\gamma_1$ and $\gamma_2$\cite{AguadoN,Aguado2,Wimmer,JieA,QF1}. 
Coupling such an ABS to an external reservoir (enabled by their finite spatial overlap) introduces two key couplings: an internal hybridization $E_1$ between $\gamma_1$ and $\gamma_2$, and an external coupling $t_1$ between $\gamma_2$ and the reservoir. 
These couplings generate additional dynamical phases and thereby spoil the non-Abelian braiding properties of MZMs\cite{JieA,QF1,cole,Rachel}. 
Recent theoretical analyses show that such unwanted dynamical effects can be suppressed provided the braiding time satisfies $T_B \ll \hbar/\sqrt{E_1^2 + t_1^2}$ \cite{QF1}. 
Therefore, the magnitudes of $E_1$ and $t_1$ directly control the fidelity of braiding operations. 
Intriguingly, under optimized conditions, ABS-based qubits may even outperform conventional MZMs in realistic devices when these couplings are sufficiently small\cite{Aguado2,Wimmer,QF1}. 
This suggests that well-engineered ABSs can serve as viable platforms for topological quantum computation. 
A notable realization is the minimal Kitaev chain\cite{MkT1,MkT2,MkT3,MkT4,Mk1,Mk2,Mk3,Mk4}, where two QDs coupled via a hybrid superconducting structure realize effectively fine-tuned ABSs. 
At appropriate ``sweet spots,'' these modes exhibit quadratic protection against global perturbations of the QD potentials\cite{Mk4}. 
Moreover, recent advances in single-shot parity readout, notably by Microsoft and the Delft group\cite{Mq4,KC}, enable direct access to MQs states. 
In light of these developments, quantifying the stability of MQs in realistic setups is more meaningful than rigidly classifying low-energy states as either MZMs or ABSs. 
Such a perspective is essential for advancing fault-tolerant topological quantum computing, as it directly connects experimentally accessible observables to the microscopic parameters that determine qubit performance.

In this work, we propose a framework to quantify MQs stability through the measurement of Rabi oscillations. Rabi oscillations, a cornerstone technique in conventional quantum control, have recently been theoretically suggested in MQs\cite{Ra1, Ra2}.
 This opens the door to applying established quantum control methodologies in topological platforms. 
 Our approach involves coupling a QD to the MQs and monitoring the induced Rabi oscillations. In an ideal system, the Rabi frequencies would be identical for all coupled states. 
 However, in realistic systems with finite-size effects and disorder, small energy splittings arise, leading to observable beating patterns in the Rabi signal. These beatings serve as a direct, quantitative probe of MQs stability.
Moreover, the oscillations directly manifest as variations in the charge occupancy of the QD, which can be detected using recently developed single-shot readout techniques. 
This suggests that our proposal can be readily implemented in current experiments. 
We further investigate the impact of quasiparticle poisoning on the proposed stability measure. 
We find that weak dissipation does not alter the beating structure, indicating robustness against local decoherence. 
Importantly, in an ideal Majorana system, dissipation acts locally and affects only the Majorana mode it couples to, without propagating to its nonlocal partner. 
As a result, the coherence encoded in the spatially separated Majorana pair remains preserved. 
This behavior reflects the intrinsic nonlocal nature of topological qubits and their resilience to local decoherence.

\begin{figure}
\centering
\includegraphics[width=3.25in]{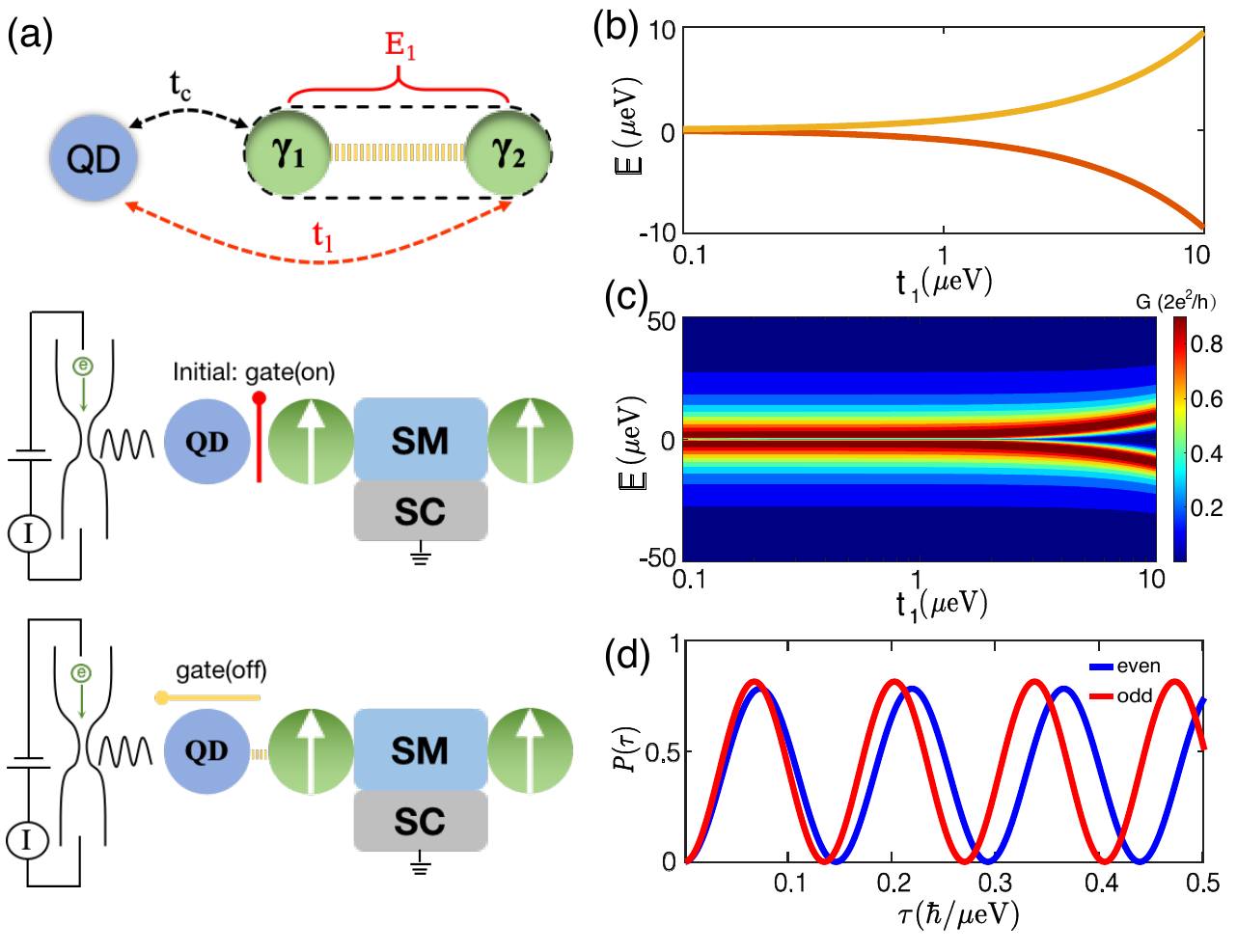}
\caption{
(a) Schematic of the QD--MQ setup. Non-ideal effects generate finite $t_1$ and $E_1$ (top). The middle and bottom panels illustrate the protocol: the QD is initially decoupled from the minimal Kitaev chain and then abruptly coupled by switching on the gate. 
(b) Near-zero-energy spectrum as a function of $t_1$, showing the deviation from zero energy with increasing $t_1$. 
(c) Corresponding conductance for (b), where small $t_1$ is difficult to resolve. 
(d) Rabi oscillations in the QD--MQ system; finite $t_1$ and $E_1$ lead to a parity-dependent frequency splitting.
}
\label{f1}
\end{figure}

\textit{Model-independent Hamiltonian and quantum master equation.} 
This setup corresponds to a QD--TS hybrid system that can be realized across multiple experimental platforms, 
including nanowire-based devices (see Fig.~\ref{f1}(a))~\cite{deng2} and vortex-based systems probed by scanning tunneling microscopy, 
in which an attached molecule serves as the QD~\cite{Jievortex}. 
This versatility provides a unified framework for assessing MQ stability across diverse architectures.
The system is generically described by the following model-independent Hamiltonian:
\begin{equation}
H_E =
E_d (1 - 2 d^\dagger d)
- i E_1 \gamma_1 \gamma_2
+ t_c (d^\dagger - d)\gamma_1
- i t_1 (d + d^\dagger)\gamma_2 .
\end{equation}
Here, $d$ represents the annihilation operator for the fermionic state in the QD, and  $E_d$ denotes the on-site energy of this QD state. 
We define MQs using the fermionic operator $f=(\gamma_1+\mathrm{i}\gamma_2)/2$, which is composed of two MZMs. 
The coupling between the QD and the MQs is characterized by $t_c$ (coupling to $\gamma_1$) and $t_1$ (coupling to $\gamma_2$). 
In addition, $E_1$ denotes the internal hybridization between $\gamma_1$ and $\gamma_2$ within the MQ. 
In realistic systems—such as ABSs induced by inhomogeneous potentials or MZMs in finite-sized wires—both $E_1$ and $t_1$ generally become finite, thereby degrading MQ stability. 
A particularly common situation arises when $E_1 \to 0$ while $t_1$ remains finite, leading to a zero-bias conductance peak. 
In this regime, coupling to a QD shifts the MQs away from zero energy [Fig.~\ref{f1}(b)]. 
While such deviations can in principle be identified from the energy spectrum \cite{AguadoN} and have been observed in impurity-induced ABSs \cite{deng2,Vf}, conductance measurements can resolve them only when $t_1$ is sufficiently large [Fig.~\ref{f1}(c)]. 
As a result, small but finite couplings—most relevant for MQ stability—remain effectively inaccessible to conventional probes, motivating the need for more sensitive detection schemes in the weak-coupling regime.

To address this limitation, we propose using Rabi oscillations to quantify MQ stability. 
To characterize their dynamics, we compute the density matrix $\rho$ by solving the Lindblad master equation:
\begin{equation}  \label{Eq2}
\dot{\rho}(\tau)=\frac{-\mathrm{i}}{\hbar}[\hat{H},\rho(\tau)]+\sum_i\Gamma_i[L_i\rho(\tau)L_i^{\dag}-\frac{1}{2}\{L_i^{\dag}L_i,\rho(\tau)\}].
\end{equation}
Here, $L_i$ are the Lindblad operators that describe the dephasing terms in the TS with strength $\Gamma_i$. 
The Hamiltonian $\hat{H}$ in Eq.~(1), written in the qubit basis 
$(|0\rangle, d^{\dagger}f^{\dagger}|0\rangle, f^{\dagger}|0\rangle, d^{\dagger}|0\rangle)$, 
block-diagonalizes into even- and odd-parity sectors of the QD--MQ system, 
$\hat{H} = H_{+} \oplus H_{-}$. 
The corresponding Hamiltonians read
\begin{equation}
H_{\pm}= \begin{bmatrix}
   E_d\pm E_1 & t_c\pm t_1 \\
   t_c\pm t_1 & -E_d\mp E_1
  \end{bmatrix}.
\end{equation}
In the ideal limit $E_1 = t_1 = 0$, the two parity sectors are identical. 
Finite $E_1$ and $t_1$ lift this degeneracy, leading to a small splitting between the corresponding Rabi frequencies. 
In this work, we focus on the weak-splitting regime, where $E_1$ and $t_1$ are small and give rise to the beating dynamics discussed below.

\begin{figure}
\centering
\includegraphics[width=3.25in]{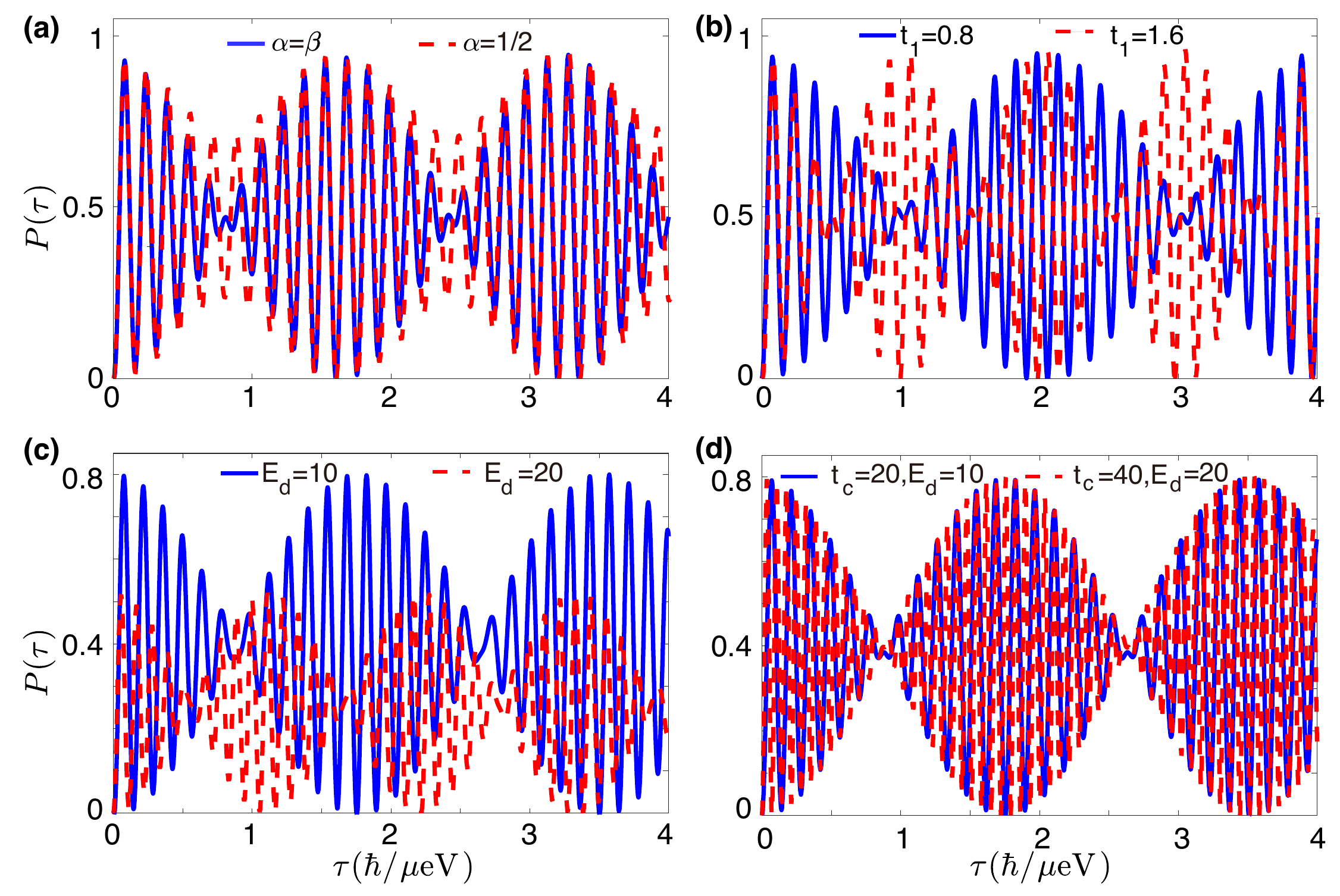}
\caption{
Typical Rabi beats. 
(a) Rabi beats in the mixed-parity regime,  Parameters: $t_c = 20\,\mu\text{eV}$, $E_d = 5\,\mu\text{eV}$, $t_1 = 1\,\mu\text{eV}$, and $E_1 = 0$.
(b) Rabi beats for $t_1 = 0.8\,\mu\text{eV}$ and $t_1 = 1.6\,\mu\text{eV}$, showing that the beating frequency scales linearly with $t_1$. Here $E_1 = 0$, and other parameters are the same as in (a).
(c) Rabi beats for $E_d = 10\,\mu\text{eV}$ and $E_d = 20\,\mu\text{eV}$ with $E_1 = 2\,\mu\text{eV}$ and $t_1 = 0$. The beating frequency scales with $\cos\theta$ in this case, in contrast to the $\sin\theta$ dependence when $t_1 \neq 0$ and $E_1 = 0$.
(d) Rabi beats for $(t_c, E_d) = (20\,\mu\text{eV}, 10\,\mu\text{eV})$ and $(40\,\mu\text{eV}, 20\,\mu\text{eV})$ with $t_1 = 0.5\,\mu\text{eV}$. The beating frequency remains unchanged despite a doubling of the base Rabi frequency $\Omega_0$.}
\label{f2}
\end{figure}

\textit{Rabi beats in MQ.} 
Rabi oscillations can be driven by voltage pulses or microwave excitation. We consider a direct protocol in which the QD is initially decoupled from the MQ [Fig.~1(a)], although an equivalent implementation via microwave driving is also possible.
 At time $t=0$, a voltage pulse switches off gate $G1$, thereby coupling the QD to the MQ. 
 This sudden quench breaks charge conservation on the QD and induces oscillations in its occupation, constituting Rabi oscillations. The resulting charge fluctuations can be directly probed via single-shot readout techniques, 
 as recently demonstrated by Microsoft and the Delft group \cite{Mq1,KC,Mq2,Mq3,Mq4}, indicating experimental feasibility. We assume the QD is initially unoccupied.
In the absence of dissipation, fermion parity is conserved, and the system decouples into two independent subspaces, $H_{+}$ and $H_{-}$. Each subspace forms an effective two-level system, and the charge variance in each sector is given by
\begin{equation} \label{Eq1}
P_{\pm}(\tau)=\frac{(t_c\pm t_1)^2}{(E_d\pm E_1)^2+(t_c\pm t_1)^2}\sin^2(\Omega_{\pm}\tau).
\end{equation}
Here $\Omega_{\pm}=\sqrt{(E_d\pm E_1)^2+(t_c\pm t_1)^2}$ are the Rabi frequencies in the even ($+$) and odd ($-$) parity sectors. For small $E_1$ and $t_1$, one finds
$\Omega_{\pm}\approx \Omega_0 \pm \left(E_1\cos(\theta) + t_1\sin(\theta)\right)$,
where $\Omega_0=\sqrt{E_d^2+t_c^2}$, and $\sin(\theta) = t_c/\Omega_0$, which characterizes the amplitude of the Rabi oscillations. In the ideal limit $E_1=t_1=0$, the two frequencies coincide, $\Omega_{\pm}=\Omega_0$. Finite $E_1$ and $t_1$ lift this degeneracy; although the difference is not apparent at short times, it accumulates during time evolution, leading to observable deviations, as shown in Fig.~1(d).

If the system occupies a superposition of parity sectors, the frequency mismatch manifests as a beating pattern in the dynamics. 
This is precisely the situation in the QD--MQ setup. Since the MQ energy is close to zero, its state can form an arbitrary qubit superposition, 
$\alpha |0\rangle + \beta f^{\dagger}|0\rangle$, while the QD is initialized in a definite charge state. 
This configuration corresponds to a superposition of even and odd parity sectors (i.e., mixed parity). 
The total charge variance is then given by
$P(\tau) = |\alpha|^2 P_{+}(\tau) + |\beta|^2 P_{-}(\tau)$. 
For the equal-weight case $|\alpha|=|\beta|=1/\sqrt{2}$, one obtains
\begin{equation} \label{Eq2}
P(\tau)=\frac{\sin^2\theta}{2}\left[1-\cos(2\Omega_0\tau)\cos(2\delta\Omega\,\tau)\right],
\end{equation}
where $\delta\Omega = E_1\cos\theta + t_1\sin\theta$. 
Both the oscillation amplitude $(\sin^2\theta)/2$ and the beating frequency $2\delta\Omega$ are directly accessible experimentally. 
Moreover, by tuning $\theta$ through system parameters (e.g., $E_d$ and $t_c$), the individual contributions of $E_1$ and $t_1$ can be disentangled, 
providing a direct and quantitative determination of the MQ deviation parameters.

Figure~2(a) shows the QD charge variance $P(\tau)$ in the mixed-parity regime, where a clear beating pattern emerges, consistent with Eq.~(\ref{Eq2}) (blue line). 
Notably, the beating persists even for unequal weights $\alpha \neq \beta$ (red dashed line), demonstrating its robustness. 
Figure~2(b) shows $P(\tau)$ for $t_1=0.8\,\mu\text{eV}$ and $t_1=1.6\,\mu\text{eV}$ at $E_1=0$. 
The beating frequency doubles as $t_1$ is doubled, demonstrating its linear dependence on $t_1$ in this limit. 
In the complementary case $t_1=0$ with finite $E_1$, the beating frequency scales with $\cos\theta$, consistent with Eq.~(\ref{Eq2}), as illustrated in Fig.~2(c). 
More importantly, the beating frequency depends only on the splitting $\delta\Omega$ and is independent of the base Rabi frequency $\Omega_0$. 
As shown in Fig.~2(d), doubling $\Omega_0$ while keeping the oscillation amplitude unchanged leaves the beating frequency unaffected, 
highlighting its robustness against variations in the overall energy scale. 
Taken together, these results establish the beating frequency as a direct, quantitative, and experimentally accessible probe of MQ stability.

\begin{figure}
\centering
\includegraphics[width=3.25in]{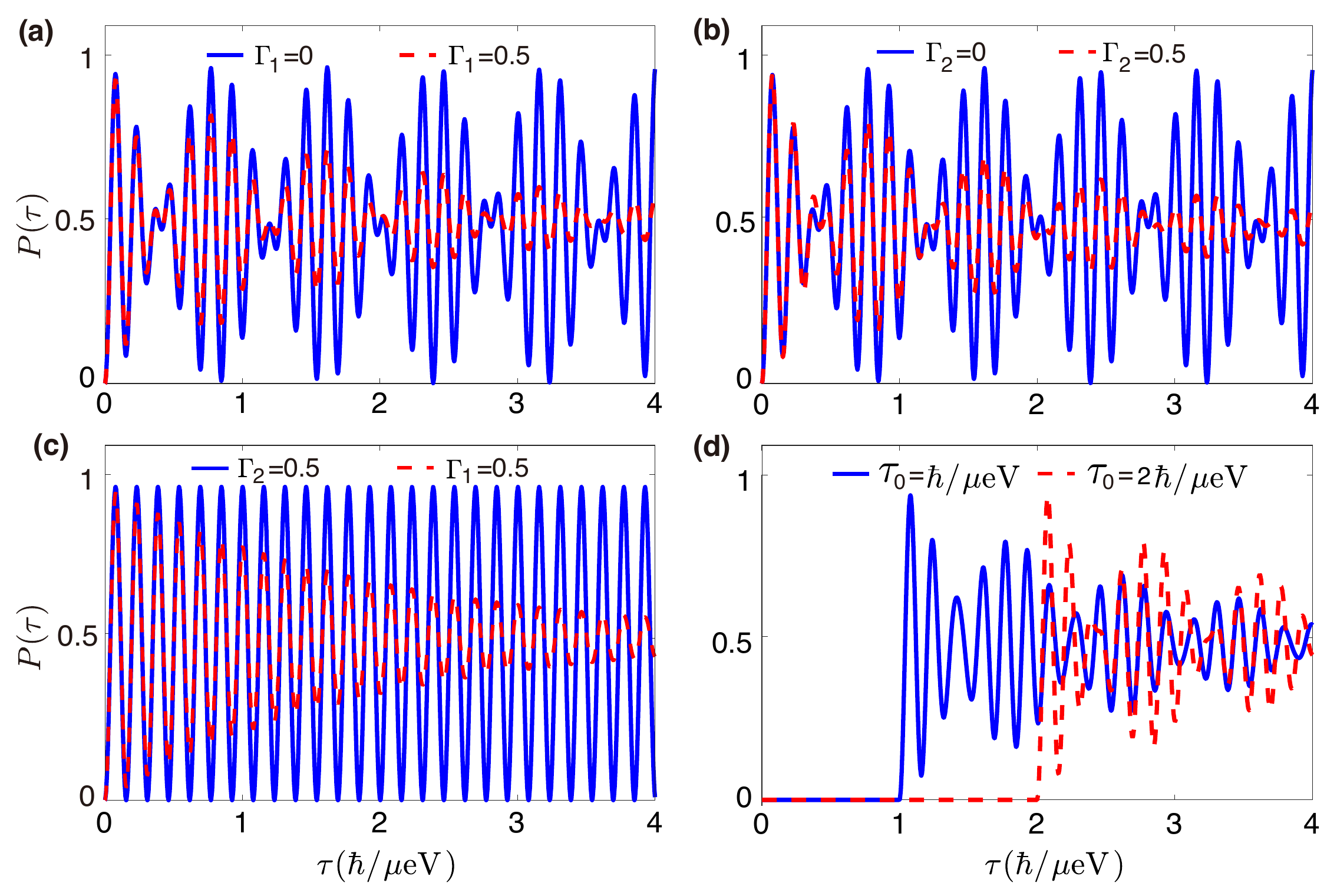}
\caption{ Dissipation effects in MQ. 
(a),(b) Influence of dissipative channels $L_1$ and $L_2$ on the Rabi dynamics. Parameters: $t_c = 20\,\mu\text{eV}$, $E_d = 5\,\mu\text{eV}$, $t_1 = 2\,\mu\text{eV}$, and $E_1 = 0$. Dissipation induces decay of the Rabi oscillations and drives the system toward mixed-parity states. 
(c) Distinct effects of $L_1$ and $L_2$ in the ideal MQ limit ($E_1 = t_1 = 0$), where dissipation acts differently on local and nonlocal degrees of freedom. Other parameters are the same as in (a).
(d) Dissipation prepares the needed mixed states  on a timescale $\sim 2\hbar/\mu\text{eV}$. A subsequent sudden coupling induces Rabi beating identical to that in (b). Parameters are the same as in (b) with $\Gamma_2 = 0.5\,\mu\text{eV}$, except that the initial state is $|0\rangle$.}
\label{f3}
\end{figure}

\textit{Influence of dissipation in MQ.} 
The emergence of Rabi beating relies on the MQ being prepared in a superposition $\alpha |0\rangle + \beta f^{\dagger}|0\rangle$. 
This raises the question of how such a state can be initialized. 
In general, dissipation tends to drive the system into mixed states while suppressing quantum coherence, thereby obscuring interference effects such as beating. 
It is therefore crucial to understand whether the beating pattern survives under realistic conditions and whether dissipation can be compatible with the preparation of coherent MQ states. 
In the following, we investigate the impact of dissipation on the Rabi dynamics and show that, under appropriate conditions, the beating pattern remains robust. 
This provides a viable route to initializing MQ states without destroying the beating effect.

We model dissipation via Lindblad operators $L_i = \sqrt{\Gamma_i}\,\gamma_i$ ($i=1,2$), 
where $\gamma_i$ are Majorana operators and $\Gamma_i$ the corresponding decay rates~\cite{Bo1}. 
Figures~\ref{f3}(a) and~\ref{f3}(b) show the effect of dissipation induced by $L_1$ and $L_2$, respectively, for $E_1 = 2\,\mu\text{eV}$. 
In both cases, the QD occupation amplitude decays over time and approaches a steady value, while the beating frequency remains unchanged, allowing it to be reliably extracted even in the presence of dissipation. 
A striking asymmetry emerges in the ideal MQ limit ($E_1 = t_1 = 0$). 
As shown in Fig.~\ref{f3}(c), dissipation through $\gamma_2$ ($\Gamma_2 = 0.5\,\mu\text{eV}$, blue solid line) leaves the Rabi oscillations essentially undamped, 
whereas the same strength of dissipation through $\gamma_1$ (red dashed line) leads to rapid decay. 
This behavior reflects the nonlocal nature of Majorana modes: in the ideal limit, the two Majoranas are spatially separated, 
and dissipation acts only locally on the Majorana it couples to. 
As a result, noise acting on $\gamma_2$ does not propagate to $\gamma_1$, and therefore does not affect the QD–MQ hybridization responsible for the Rabi dynamics. 
In contrast, $\gamma_1$ is directly involved in the hybridization with the QD, so dissipation through $L_1$ leads to decoherence and damping.
Consequently, dissipation in an ideal MQ affects only local degrees of freedom and does not destroy the nonlocal coherence encoded in the Majorana pair. 
This locality ensures that the beating frequency remains robust against weak dissipation. 
Moreover, dissipation can naturally drive the system toward a mixed-parity configuration, providing a practical route to initializing the states required for observing the beating dynamics.
A practical strategy to initialize such a mixed state is to introduce weak dissipation on $\gamma_2$ (i.e., at the right end of the Majorana wires). 
In the weak deviation case, this dissipative channel drives the MQ into the steady state $(|0\rangle + f^{\dagger}|0\rangle)/\sqrt{2}$, as illustrated in Fig.~\ref{f3}(d). 
Starting from $|0\rangle$ and adding only $L_2$ dissipation, the system relaxes to the maximally coherent superposition after a sufficiently long dwell time $\tau_0$. 
Subsequently applying a voltage pulse to couple the QD yields Rabi beats that are essentially unaffected by the same dissipation. 
Hence, the beating pattern remains observable and dissipation itself can be harnessed for deterministic MQ state preparation.
 
\begin{figure}
\centering
\includegraphics[width=3.25in]{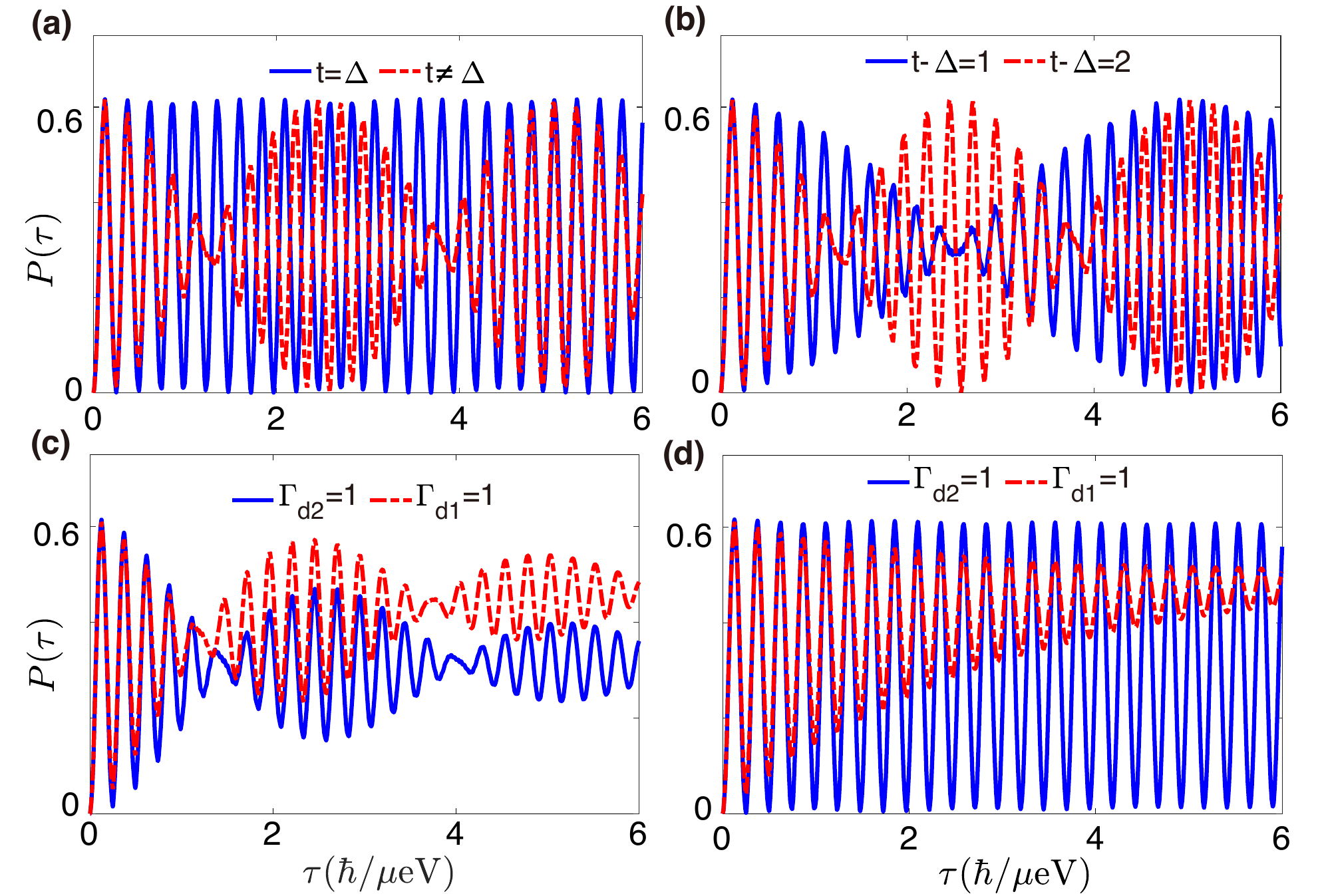}
\caption{ Rabi oscillation in minimal kitaev chain. Here we fixed $t=100\,\mu\text{eV}$,  $t_d=20\,\mu\text{eV}$, $E_d=8\,\mu\text{eV}$.
(a) Rabi oscillation for $\Delta=100\,\mu\text{eV}$ (blue solid line) and $\Delta=102\,\mu\text{eV}$ (red dashed line).
(b) Beating frequency is proportional to the deviation of $t-\Delta$. 
(c)  Rabi oscillation in the influence of $L_1=d_1$ (red line) and $L_2 =d_2$ (blue line) at $t-\Delta=2\,\mu\text{eV}$.
 (d) Rabi oscillation in the influence of $L_1=d_1$ (red line) and $L_2 =d_2$ (blue line) at $t=\Delta$.}
\label{f4}
\end{figure}

\textit{Rabi Oscillation  in minimal Kitaev chain.}  Finally, we investigate the Rabi oscillations in a real system—specifically, a minimal Kitaev chain. 
In this case, we assume that the chemical potential of the quantum dots in the chain is well-controlled and remains at zero energy. 
Therefore, the Hamiltonian of the minimal Kitaev chain is given by: $H_M = td_1^{\dag}d_2+\Delta d_1d_2+h.c.$. 
Here, \( t \) represents the hopping term between the two QDs, and \( \Delta \) is the pairing term between the two QDs. 
When the system is at the "sweet spot", where \( t = \Delta \), two MZMs will exist in the QDs, respectively. 
Now, we assume that another QD is attached to the chain. The total Hamiltonian for this system is then given by:
$H_T = -E_d d^{\dagger} d+t_d d^{\dagger}d_1+h.c. +H_M$, where $t_d$ is the hopping term between the QD and Kitaev chain.

Following the previous proposal, we set the initial state of QD is unoccupied while the initial state of minimal Kitaev chain is a superposition of even parity and odd parity.
 In this situation, we can see that the results are in full consistent with the effective model. The Rabi beats can emerge in the situation that \( t \neq \Delta \) (see the corresponding red line in Fig. 4(a)). 
 Furthermore, the Rabi frequency is proportional to the deviation of $t-\Delta$ (See Fig. 4(b)). This suggests that Rabi beat indeed can be used to quantify the stability of MQ. 
 We further investigate the influence of dissipation in this minimal Kitaev chain. The dissipation is modeled by Lindblad operators $L_i = d_i$ with rates $\Gamma_{d_i}$, which describe particle loss processes in QD1 and QD2. 
In the non-ideal case ($t \neq \Delta$), dissipation in either QD1 or QD2 leads to a suppression of the Rabi oscillation amplitude, as shown in Fig.~4(c). 
In contrast, at the sweet spot ($t = \Delta$), dissipation acting on QD1 and QD2 has qualitatively different effects. 
All these features are well captured by the effective model, demonstrating that it accurately describes the essential physics of the Rabi dynamics in the MQ system.

\textit{Conclusion.} 
We have shown that coupling a QD to an MQ enables direct access to its Rabi dynamics. 
In realistic systems, finite-size effects and inhomogeneous potentials give rise to Rabi beating, with a frequency $2\delta\Omega = 2(E_1\cos\theta + t_1\sin\theta)$ set by the MQ deviation parameters $E_1$ and $t_1$. 
Since all other parameters can be independently extracted from experimental measurements, the beating pattern provides a direct and quantitative route to determine these deviations. 
We further investigate the role of dissipation and find that the beating frequency remains robust against weak disspative processes, which primarily lead to a decay of the oscillation amplitude without altering the underlying frequency splitting. 
Moreover, dissipation can assist in driving the system toward a mixed-parity configuration, providing a practical route for initializing the states required to observe the beating pattern. 
Our results establish a practical and experimentally accessible approach to characterizing MQ stability and provide new insight into the dynamical consequences of Majorana nonlocality.

Finally, we address several considerations relevant to realistic implementations. 
First, Rabi protocols often assume sudden coupling, which may induce diabatic effects. To avoid this, the driving strength should remain below the superconducting gap. 
In practice, microwave driving typically operates at strengths of a few $\mu\text{eV}$, well below the superconducting gap (of order meV), ensuring the validity of this condition\cite{Kim}.
Second, the beating frequency directly reflects the MQ deviations. For deviations on the order of a few $\mu\text{eV}$, the corresponding timescale is on the order of nanoseconds, well within the resolution of QD charge qubit measurements. 
For smaller deviations $\sim 0.1\,\mu\text{eV}$, the timescale extends to tens of nanoseconds, further relaxing the temporal resolution requirements and enhancing experimental accessibility.
Finally, while Rabi oscillations provide a basic mechanism for probing MQ dynamics, more advanced protocols, such as Ramsey interferometry, may offer enhanced sensitivity, suggesting a promising direction for further improving the characterization of MQ stability.

\textit{Acknowledgement.} Jie Liu acknowledge the fruitful discussion with Ji‐Yin Wang. This work is financially supported by National Natural Science Foundation of China (Grants No. 11974271, and No. 92265103) , 
 the Innovation Program for Quantum Science and Technology (Grant No. 2021ZD0302400), and Shaanxi Fundamental Science Research Project for Mathematics and Physics (Grant No.25JSY003).

\end{document}